\documentclass[twocolumn,showpacs,pre]{revtex4}

\usepackage{graphicx}

\def\bea{\begin{eqnarray}}
\def\eea{\end{eqnarray}}
\def\D{\Delta}

\def\d{\delta}
\def\p{\partial} 
\def\nn{\nonumber}
\def\up{\uparrow}
\def\dn{\downarrow}

\def\la{\langle}
\def\ra{\rangle}
\def\th{\epsilon}

\def\t{\tau}

\def\f{\frac}

\def\s{\sigma}

\begin{document}

\title{ Work distribution functions for hysteresis loops in a
  single-spin system }
\author{Rahul Marathe and Abhishek Dhar}
\email{dabhi@rri.res.in}
\affiliation{Raman Research Institute, Bangalore 560080}
\date{\today} 

\begin{abstract}
We compute the distribution of the work done in driving a single Ising 
spin with a time-dependent magnetic field. 
Using Glauber dynamics we perform Monte-Carlo simulations to find the
work distributions at different driving rates. We find that in general
the work-distributions are broad with a significant probability for
processes with negative dissipated work . The special cases of slow and  fast driving
rates are studied analytically. We verify that various work
fluctuation theorems corresponding to equilibrium initial states are
satisfied while a steady state version is not. 
\end{abstract}

\maketitle

\section{Introduction}

Consider a magnetic system in a time-dependent magnetic 
field. Assume that  magnetic field is varied periodically. Then plotting the
magnetization of the system against the instantaneous magnetic 
field we get the well-known hysteresis curve. The area enclosed by the
hysteresis loop gives the work done on the system by the external
field and this leads to heating of the magnet. In the usual picture
that one has of hysteresis  one expects that the work done is
positive. However if the magnetic system is small ({\emph i.e}
contains small number of magnetic moments) then  this is no longer
true. For a small magnet one finds that the work is a fluctuating
quantity and in a particular realization of the hysteresis experiment
one could actually find that the magnet cools and does work on the
driving force.

In general, for a {\emph small system} driven by time-dependent forces whose
rates are not slow compared to relaxation times, one typically
finds that various nonequilibrium quantities,  such as the work done or
the heat exchanged, take values from a distribution. 
Recently there has been a lot of interest in the properties of such
distributions.  Part of the reason for the interest is that it leads
us to examine the question as to how the usual laws of thermodynamics,
which are true for macroscopic systems, need to be modified when we
deal with mesoscopic systems \cite{phystod}.

For instance in our example of the magnet with a small number of spins,
there is a finite probability that all the spins could suddenly 
spontaneously flip against the direction of the field by drawing
energy from the heat bath.  
Intuitively this gives one the feeling that there has been a
violation of the second law. In fact historically early observers of
Brownian motion had the same feeling when they saw the ``perpetual''
motion of the Brownian particles \cite{perrin}. However if one looks at the precise
statement of the second law one realizes that there is no real
violation.The second law is a statement on the most probable
behavior while here we are looking at fluctuations about the most
probable values. These  become extremely small for thermodynamic
systems. On the other hand, for small systems these fluctuations are
significant and a study of the properties of these fluctuations could
provide us with a better understanding of the meaning of the second law in the
present context. 
This will be necessary for an understanding of the behavior of
mesoscopic systems such as molecular motors, nanomagnets, quantum dots
etc. which are currently areas of active experimental interest.

Much of the recent interest on these nonequilibrium fluctuations has
focused on two interesting results on the distribution of the
fluctuations. These are (1) the Jarzynski relation \cite{jarz,cohen} and (2) the
fluctuation theorems \cite{deter,crooks,stat,onut,cohen2}. A large number
of studies, both theoretical and experimental \cite{lip,wang,expt} have looked at the
validity of these theorems  in a variety of systems and also their
implications. At a fundamental level both these theorems give some
measure of ``second law violations''.  At a practical level 
the possibility of using these theorems to determine the
equilibrium free energy profile of systems using data from
nonequilibrium experiments and simulations has been explored \cite{appli}.

In this paper we will be interested in the fluctuations of the area
under a hysteresis loop for a small magnet. We look at the
simplest example, namely a single Ising spin 
in a time-dependent magnetic field and evolving through Glauber
dynamics. Hysteresis in kinetic Ising systems have been studied earlier
\cite{hyst,novo} where the main aim was to understand various
features such as dependence of the average loop area  on sweeping
rates and amplitudes, system size effects and dynamical phase
transitions . The area distribution was also studied in \cite{novo}
but the emphasis was on different aspects and so is quite incomplete
from our present viewpoint. 
A two state model with Markovian 
dynamics was earlier studied by Ritort et al \cite{ritort} to analyze experiments on
stretching of single molecules. Systems with more than two states have
also been studied \cite{peliti} in the context of single-molecule
experiments. However the detailed form of the work-distributions have not
been investigated and that is the main aim of this paper.
These distributions are of interest since there are only a few
examples where the explicit forms of the distributions have actually been
worked out \cite{speck,ritort,dhar,peliti}. Most of the experiments, for
example those in the RNA stretching experiments of Liphardt et al
\cite{lip} or
the more recent experiment of Douarche et al \cite{dour} on torsionally driven
mirrors, are in regimes where the work-distributions are Gaussian.

We perform Monte-Carlo simulations to obtain the distributions for
different driving rates. We consider different driving protocols and
look at  the two cases corresponding to the transient and the steady
state fluctuation theorems. It is shown that the limiting cases of
slow and fast driving rates can be solved analytically. We also point
out that the problem of computing work-distributions is similar to that
of computing residence-time distributions.

The organization of the paper is the following: In section (II) we
define the dynamics and derive the Fokker-Planck equation satisfied by
the distribution function. In section (III) we give the results from
Monte-Carlo simulations and in section (IV) we treat the special
cases of slow and fast driving rates. 
Finally in section (V) we give a discussion of the results.

\section{Definition of model and dynamics}
Consider a single spin, with magnetic moment $\mu$, in a time-dependent
magnetic field $h(t)$.  
The Hamiltonian is given by
\bea
H=-\mu  h  \s~~~~~~ \s=\pm 1
\eea  
We assume that the time-evolution of the spin is given by the Glauber
dynamics. Let us first consider a discretized version of the dynamics.
Let the value of the magnetic field at the end of the $(n-1)$th time
step be $h_{n-1}$ and let the value of the spin be $\s_{n-1}$. The
discrete dynamics consists of two distinct process during the $n$th time step:

1. The field is changed from $h_{n-1}$ to $h_{n}=h_{n-1}+\Delta h_n$. During
this step an amount of work $\D W= -\mu \s_{n-1} \D h_n $ is done on
the system.

2. The spin flips with probability $p (e^{-\beta \mu h_n \s_{n-1}}/Z)$
   where $Z=e^{\beta\mu h_n}+e^{-\beta\mu h_n}$ is the equilibrium
   partition function at the instantaneous field value. The factor $p$
   is a parameter that is required when we take the continuum time
   limit and whose value will set equilibration times. 
At the end of this step the spin is in the state  $\s_{n}$. During this
step the system takes in an amount of heat $\D Q= -\mu h_n
(\s_{n}-\s_{n-1})$ from the reservoir.  

Given the microscopic dynamics we can derive time-evolution equations for
various probability distributions. These are standard results but we
reproduce them here for completeness. 

{\emph{Fokker-Planck equation for spin distribution}}: First let us consider the spin
configuration probability  $P_ {n}(\s)$ which gives the probability
that at time $n$ the spin is in the state $\s$. 
We write the field in the form $h_n= h_0 f_n$ where $f_n$ is
dimensionless and let us define $\th = \beta \mu h_0$. 
Then we get the following evolution equation: 
\bea
&& \left(\begin{array}{cc}
 P_{n+1}(\uparrow)\\
P_{n+1}(\downarrow)\\
\end{array}\right)=
\left(\begin{array}{cc}
1-p \f{e^{-\th f_n}}{Z} & p \f{e^{\th f_n}}{Z} \\  
p \f{e^{-\th f_n}}{Z}& 1-p \f{e^{\th f_n}}{Z} \\
\end{array}\right)
\left(\begin{array}{cc}  
P_{n}(\uparrow)\\  
P_{n}(\downarrow)\\
\end{array}\right) \nn
\eea
To go to the continuum-time limit we  take the limits $p{\rightarrow}0$, $\D t{\rightarrow}0$,
with ${p}/{\D t} \to r$ and $f_n \to f(t)$, $P_n(\s) \to P(\s,t)$. 
Using the dimensionless time $\t=r t$ we then get:  
\bea
&&\frac{\partial \hat{P}}{\partial \t} =-{\mathcal T}\hat{P}~~~~~~{\rm where}  \\
&&\hat{P} =\left(\begin{array}{cc}  
P(\up, \t)\\  
P(\dn, \t)\\
\end{array}\right), ~~~~
{\mathcal T}= \left(\begin{array}{cc}
\f{e^{-\th f(\t)}}{Z}\ & -\f{e^{\th f(\t)}}{Z}\ \\  
- \f{e^{-\th f(\t)}}{Z}\ &  \f{e^{\th f(\t)}}{Z}\ \\
\end{array}\right). \nn
\eea
The magnetization $m(\t)=2P(\up,\t)-1$ thus satisfies the equation
\bea
\f{d m(\t)}{d \t} = -m(\t)+tanh[\th f(\t)] 
\label{mageq}
\eea
whose solution  is
\bea
m(\t)=e^{-\t} m(0) + \int_{0}^{\t}d\t^{\prime}
e^{-(\t-\t')} \tanh [\th f(\t')]. 
\eea

{\emph{Fokker-Planck equation for Work distribution}}:
The total work done at the end of the $n$th time step is given by:
\bea
W=-\mu \sum_{n}\s_{n-1}\Delta h_{n} 
\eea
To write evolution equations for the work-distribution it is necessary
to first define $Q_n(W,\s)$, the joint probability that at the end of the $n$th
step the spin is in the state $\s$ and the total work done on it is
$W$. Then $Q_n(W,\s)$ will satisfy the following recursions: 
\bea
Q_{n+1}(W,\up)&=&(1-p \f{e^{- \th f_{n+1} }}{Z_{n+1}})) Q_{n}( W + \mu
\D  h_n,\uparrow) \nn \\ &+& 
p (\f{e^{\th f_{n+1}}}{Z_{n+1}}) Q_{n}(W-\mu \D h_n,\downarrow) \nn \\
Q_{n+1}(W,\dn)&=&p (\f{e^{-\th f_{n+1} }}{Z_{n+1}}) Q_{n}( W + \mu \D
h_n,\uparrow) \nn \\ &+& 
(1-p(\f{e^{\th f_{n+1}}}{Z_{n+1}})) Q_{n}(W-\mu \D h_n,\downarrow). \nn
\eea
Taking limits $\D t \to 0$, $p \to 0$ with $p/(\D t) \to r$, $h_n \to
h(t)$ and $\D h_n / (\D t) \to \dot{h}$ and 
using dimensionless
variables $\t$ and $w=\beta W= -\th \int_0^{\tau_m} d \tau
m(\tau) df/d\tau  $ we finally get 
\bea
\frac{\partial \hat{Q}}{\partial \t} &=& -{\mathcal{ T}}. \hat{Q} +
{\th}\f{df}{d\t} {\sigma}_z. \frac{\partial \hat{Q}}{\partial w}
~~~~~~{\rm where} \label{eqQ} \\
\hat{Q} &=&\left(\begin{array}{cc}  
Q(w,\up,\t)\\  
Q(w,\dn,\t)\\
\end{array}\right), ~~
\s_z =\left(\begin{array}{cc}
 1 & 0 \\  
0  &  -1 \\
\end{array}\right) \nn 
\eea
From Eq.~\ref{eqQ} we get the following equation for
$Q(w,\t)=Q(w,\up,\t)+Q(w,\dn,\t)$: 
\bea
\f{\p^2 Q}{\p \t^2} +(1-\f{\ddot{f}}{\dot{f}})\f{\p Q}{\p \t}=
\th \dot{f} \tanh (\th f) \f{\p Q}{\p w} + (\th \dot{f})^2 \f{\p^2 Q}{\p
  w^2}
\eea  
We have not been able to solve these equations analytically except in
the limiting cases where the rate of change of the magnetic field is
very slow or very fast. In the next two sections we will first present
results from Monte-Carlo simulations which give accurate results for
any rates and then discuss the special cases. 

\section{Results from Monte-Carlo simulations}
We have studied three different driving processes: 

(A) The system is initially in equilibrium at zero field and the field
($\beta \mu h$) is 
then increased linearly from $0$ to $\th$ in time $\t_m$. 

(B) The system is initially in equilibrium and the field, which is
taken to be piecewise linear, is changed 
over one cycle.

(C) The system is  run through many cycles till
it reaches a nonequilibrium  steady state. We measure work
fluctuations in this steady state. 

In cases (A) and (B) we will be interested in testing the transient
fluctuation theorem (TFT) while in case (C) we will look at the steady state
fluctuation theorem (SSFT). Let us briefly recall the statements of
these theorems for work distributions in systems with Markovian dynamics.  

{\emph {Crook's Fluctuation Theorem}}: In this case the system is
initially in thermal equilibrium and then an external parameter [e.g magnetic
field $h(t)$] is changed from an initial value $h_i$, at time $t=0$, to a final value
$h_f$ in a finite time $t_m$. Suppose the work done on the system
during this process is $W$ and the change in equilibrium free
energy is $\Delta F$. From thermodynamics we expect that for  a
quasistatic process  $W= \Delta F$. Hence for our nonequilibrium
process it is natural to define the dissipated work as $W_d=
W-\Delta F$ which has a distribution $Q(W_d)$. Now consider a
time-reversed path for the external field $h_R(t)=h(t_m-t)$ for which
the work distribution is $Q_R(W_d)$.  
The fluctuation theorem of Crook's then states:
\bea
\f{Q(W_d)}{Q_R(-W_d)}= e^{\beta W_d}.
\label{crflth}
\eea
For Gaussian processes it can be shown that $Q_R(W_d)=Q(W_d)$
\cite{onut} and hence we get the usual form of the transient fluctuation
theorem (TFT)
\bea
\f{Q(W_d)}{Q(-W_d)}= e^{\beta W_d}.
\label{tfth}
\eea
Another situation where TFT is satisfied is the case where the field
is kept constant or if the process is time-reversal symmetric.
Finally we note that the Jarzynski relation
\bea
\la e^{-\beta  W_d} \ra = 1
\label{jeq}
\eea
follows immediately from Crook's theorem Eq.~\ref{crflth} and so will be
satisfied in all cases where Crook's holds. 

In the following sections we will verify that Crook's FT is
always satisfied but that TFT does not hold whenever the distributions are
non-Gaussian processes and the process does not have time-reversal
symmetry. We will also test the validity  of the steady state fluctuation
theorem (SSFT). This has the same form as Eq.~\ref{tfth} with the
difference that the initial state is chosen from a nonequilibrium
steady state distribution instead of an  equilibrium 
distribution.

\subsection{Field increased linearly from $0$ to $\th$}
\begin{figure}
\vspace{0.6cm}
\includegraphics[width=3.3in]{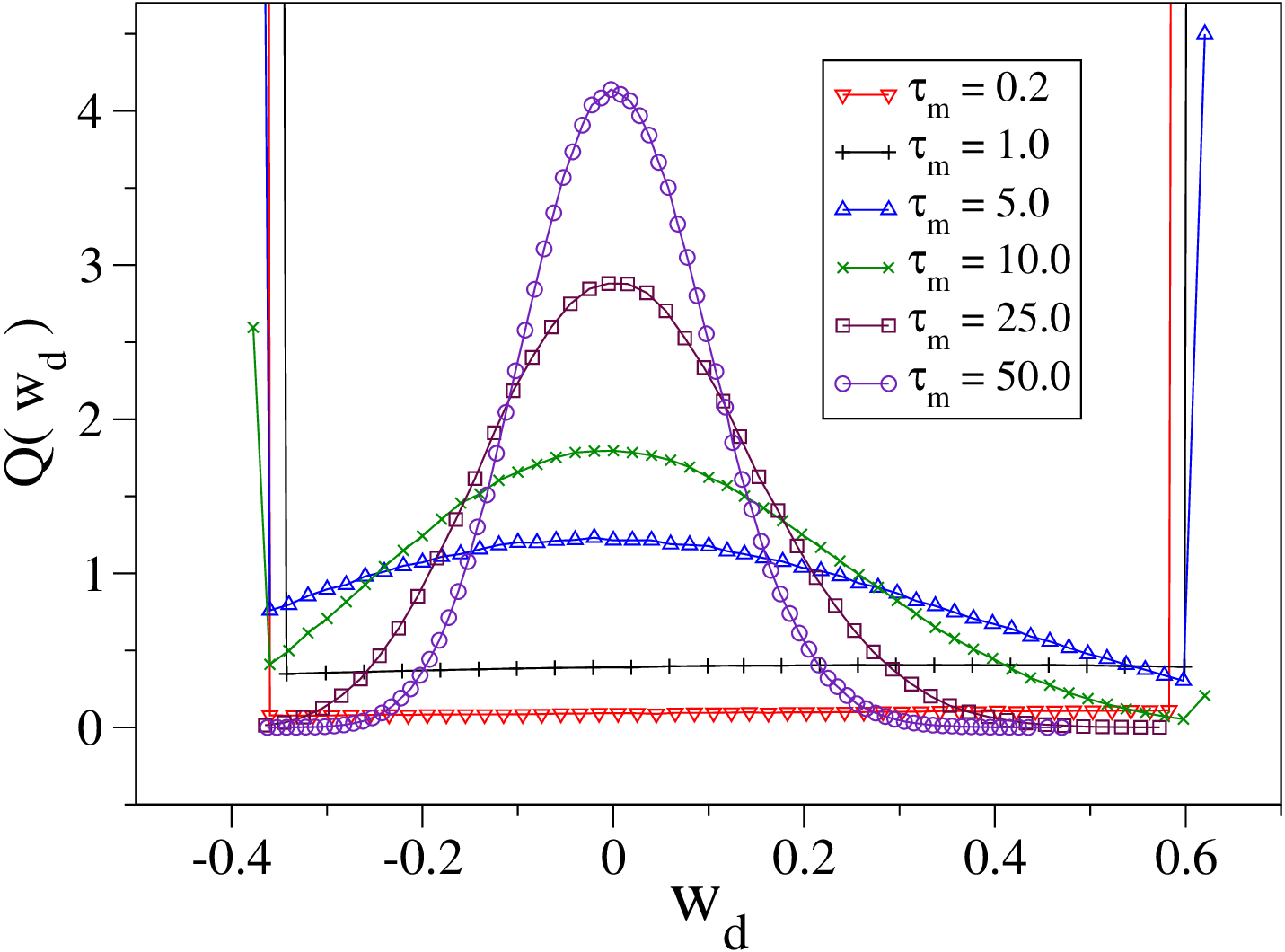}
\caption{Distributions of the work done in driving a magnet at different rates when magnetic field is changed linearly.}
\label{wdA}
\vspace{0.5cm}
\end{figure}

\begin{figure}
\vspace{0.6cm}
\includegraphics[width=3.3in]{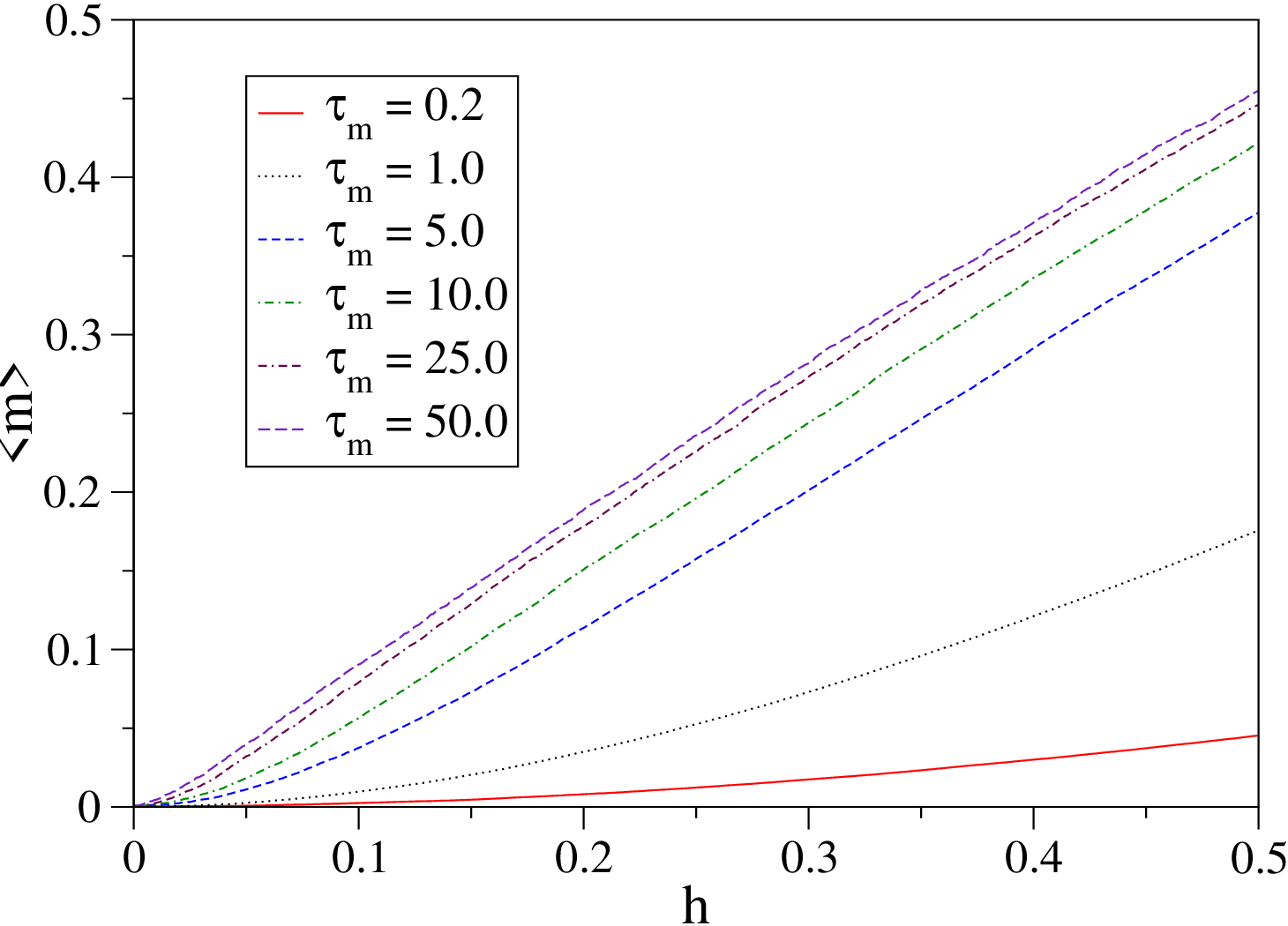}
\caption{Average magnetization $m(\t)$ for different rates, when
  magnetic field is changed linearly.} 
\label{magA} 
\end{figure}

In this case $f(\t)=\t/\t_m$ and we have chosen $\th=0.5$. We note that with a static field,
the equilibration time is given by $t_r=1/r$ or $\t_r=1$. The rate of
change of magnetic field is $1/\t_m$ and slow and fast rates correspond
 to large and small values for $\t_m$ respectively. In Fig.~\ref{wdA}
 we plot the work distributions for various values of $\t_m$. We have
 plotted the distribution of the dissipated work $w_d=w-\beta \Delta
 F$ (Here $\beta \Delta F= -\ln{\cosh{\th}}$).
  In Fig.~\ref{magA} we plot the average
 magnetization as a function of field, again for different rates. Some interesting features of
 the work-distributions  are:

(i) The distributions are in general broad. This is true  even at the
 slowest driving rates where the average magnetization
 (Fig.~\ref{magA}) itself is close to the equilibrium prediction. Note
 that the allowed range of values of $w_d$ is $[-\th-\beta\Delta
 F,\th-\beta \Delta F]\approx [-0.38,0.62]$. Also we see that the
 probability of negative dissipated work is significant.   

(ii) For slow rates the distributions are Gaussian and this can be
 understood in the following way. Imagine dividing the time range into
 small intervals. Because the rate is slow, there are a large number
 of spin flips within  each such interval, and so the average
 magnetization from one interval to the next can be expected to be
 uncorrelated. Since the work is a weighted sum of the magnetization
 over all the time intervals we can expect it to be a Gaussian. 

(iii) For fast rates we get $\d-$function peaks at $w=\pm \th$. This again is easy
 to understand since the spin doesn't have time to react and stays in
 its initial state. In section (IV) we will work out analytic
 expressions for the the work distributions by considering 
 probabilities of $0-$spin flip and  $1-$spin flip processes.  

\begin{figure}
\vspace{0.6cm}
\includegraphics[width=3.3in]{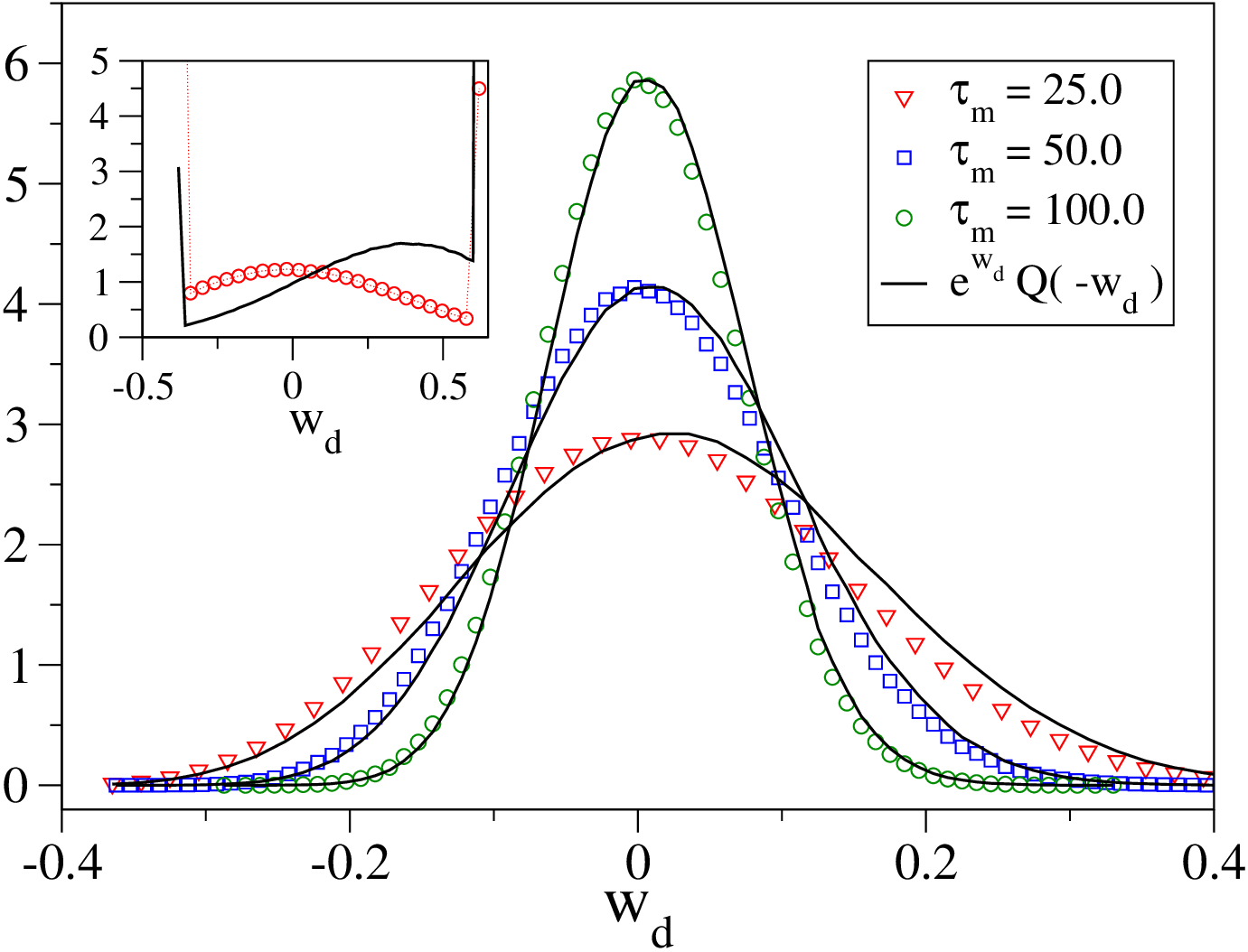}
\caption{ Plot shows that the fluctuation theorem is valid for slow
  processes with Gaussian work-distributions. Inset shows that for a
  fast rate ($\t_m=5.0$) the probability of negative work is much
  larger than that predicted by the FT.}
\label{flthA}
\vspace{0.6cm}
\end{figure}

For slow rates we have verified (see Fig.~\ref{flthA}) that the fluctuation theorem is
satisfied. For faster rates we see that the probability of
negative work processes is higher than what is predicted by the 
TFT. 
\subsection{Field is taken around a cycle}
\begin{figure}
\vspace{0.6cm}
\includegraphics[width=3.4in]{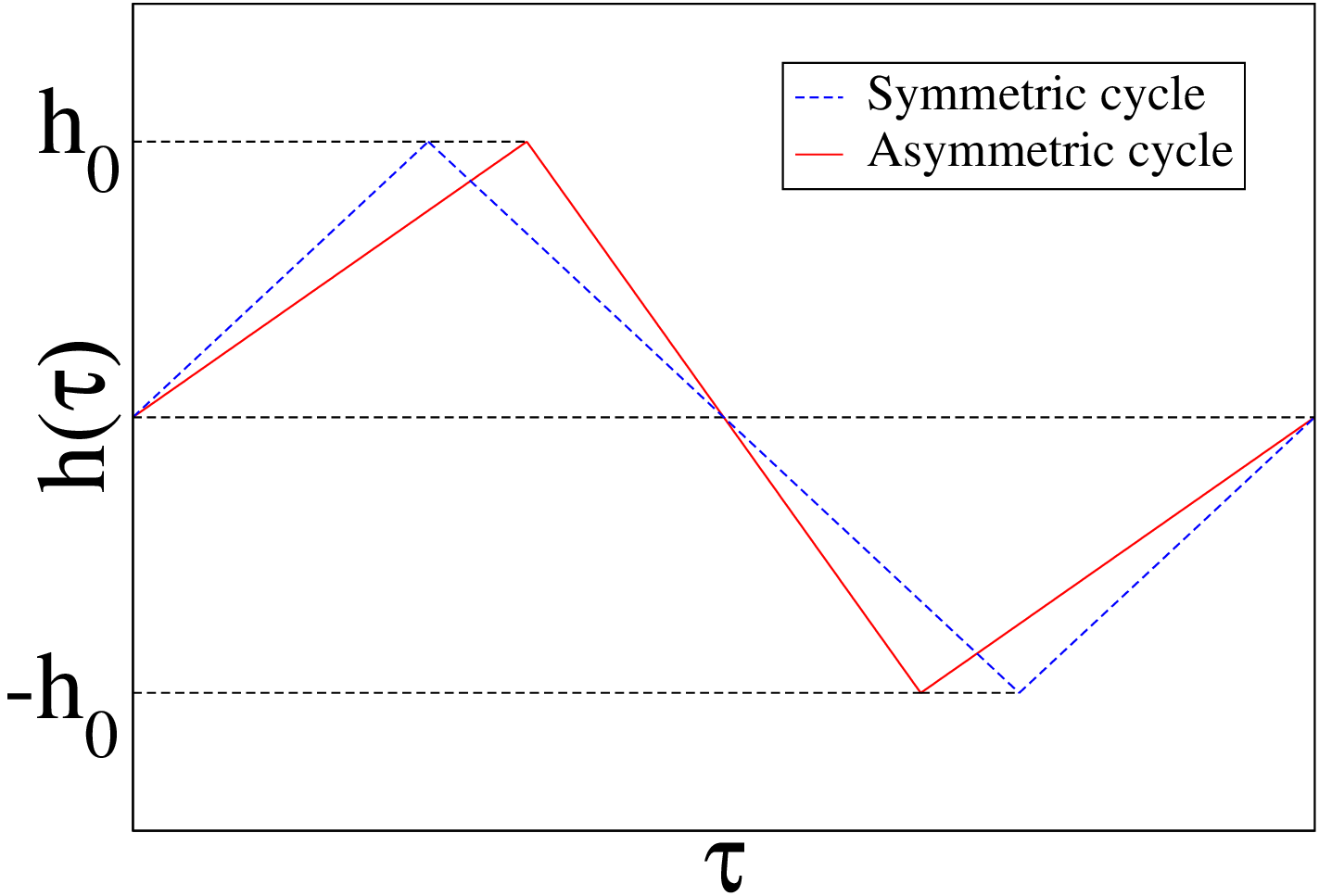}
\caption{Magnetic field changed over a cycle. In the symmetric case
  the total cycle time is $4\t_m$ while in the asymmetric case it is $2(\t_m+\t_n)$.
}
\label{saw} 
\vspace{0.65cm}
\end{figure}
\begin{figure}
\vspace{0.6cm}
\includegraphics[width=3.4in]{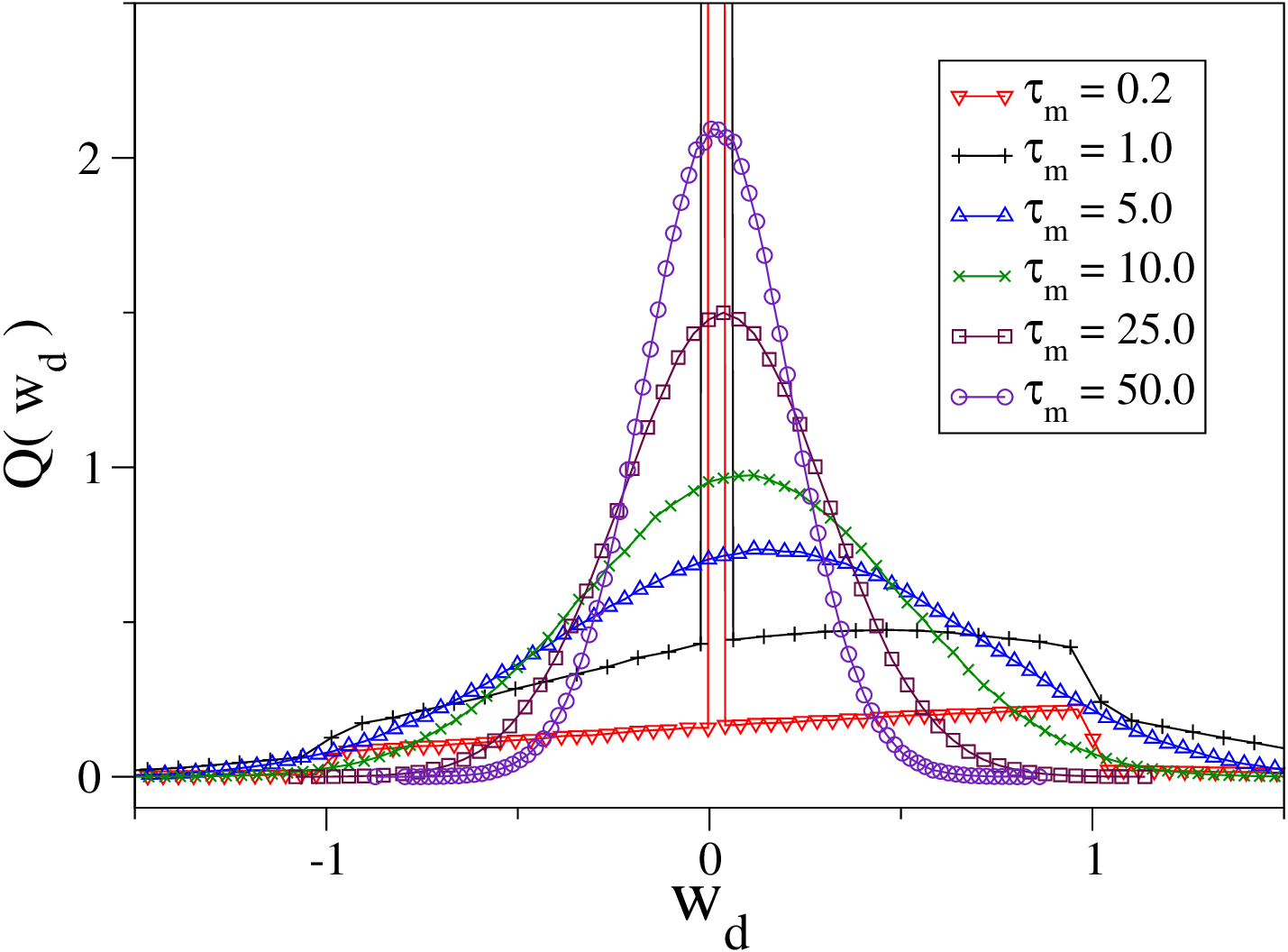}
\caption{Plot of work-distributions  for different driving rates when
  magnetic field is changed in a symmetric cycle.} 
\label{wdB1}
\vspace{0.55cm}
\end{figure}
\begin{figure}
\vspace{0.6cm}
\includegraphics[width=3.3in]{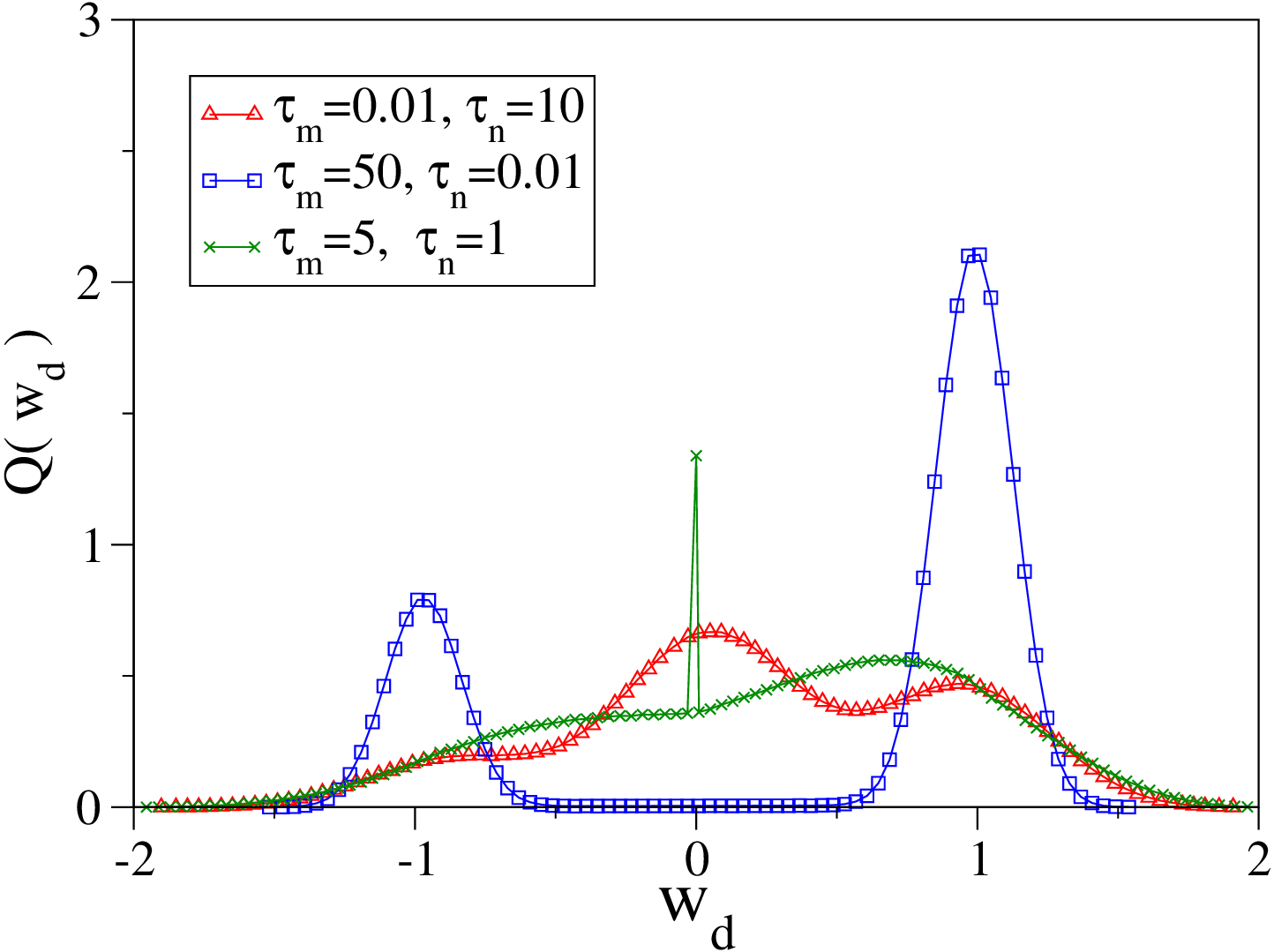}
\caption{Plot of work-distributions  obtained  for different
  asymmetric cycles of the magnetic field.} 
\label{wdB2}
\end{figure}
\begin{figure}
\vspace{0.6cm}
\includegraphics[width=3.4in]{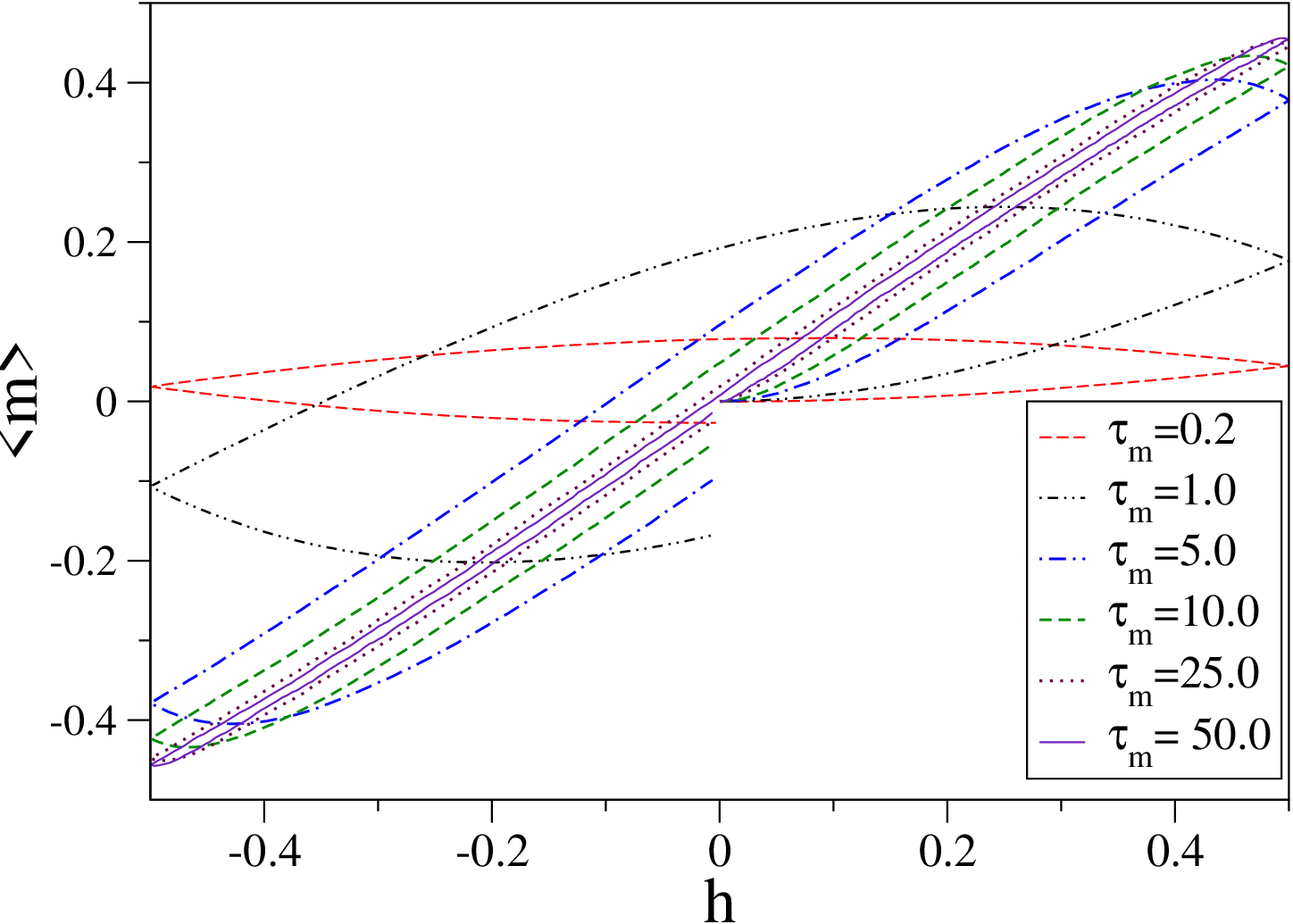}
\caption{Hysteresis curves in driving a magnet at different rates when
  magnetic field is changed in a symmetric cycle and with the spin
  initially in equilibrium.} 
\label{magB1}
\vspace{0.65cm}
\end{figure}

As shown in Fig.~\ref{saw} we consider two different cyclic forms for
$f(\t)$. One is a symmetric cycle and the other a asymmetric one. 
For these two cases the work-distributions are plotted in
Fig.~\ref{wdB1} and Fig.~\ref{wdB2} respectively. 
For the symmetric cycle we  plot the average magnetization as a function of the
field in Fig.~\ref{magB1}. This gives the familiar hysteresis curves. 

As before we again find that the work-distributions are 
broad. For slow rates we  get Gaussian distributions while for fast rates we
get a $\delta-$function peak at the origin  which correspond to a $0-$spin flip
process.  The slow and fast cases are treated analytically in
section~(III). 

As expected we can verify the transient fluctuation theorem
for both the symmetric and asymmetric processes. That TFT should be
satisfied follows from Crooks FT and noting that the time
reversed process has the same distribution as the forward process because 
of the additional $h \to -h$ symmetry that we have in this
case. 

\begin{figure}
\vspace{0.6cm}
\includegraphics[width=3.3in]{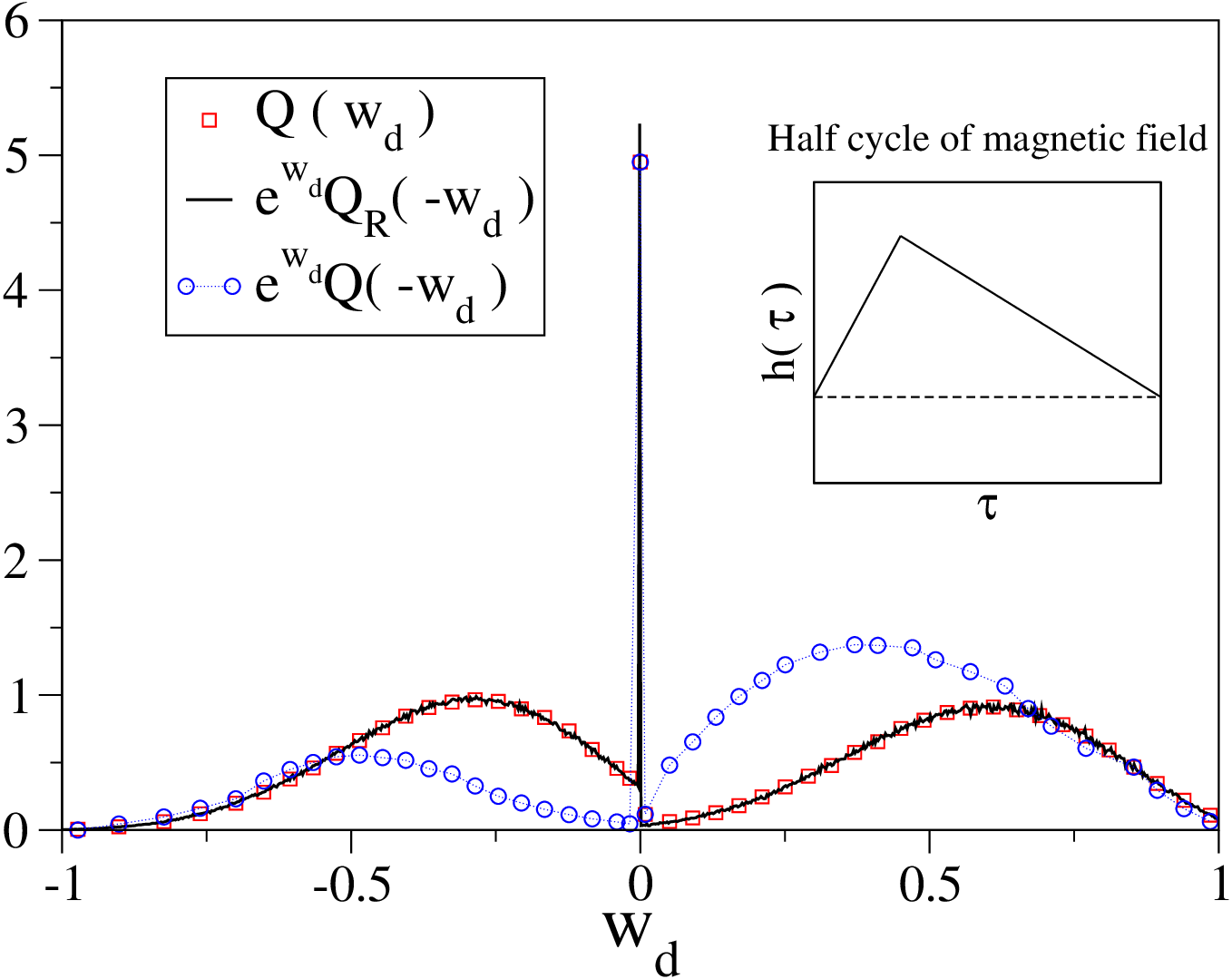}
\caption{ Plot showing the work-distribution for an asymmetric
  half-cycle and the validity of Crook's fluctuation theorem. Note that
  the probability of negative work processes is much higher than that
  predicted by usual TFT.}
\label{halfft}
\vspace{0.6cm}
\end{figure}

We have also studied an asymmetric {\emph{ half-cycle}} for which
$Q_R(w_d) \neq Q(w_d)$. 
Consequently we find that the usual TFT is not
satisfied while the more general form of TFT of Crooks holds.    
We show this in Fig.~\ref{halfft} where we have plotted
$Q(w_d)$, $e^{w_d} Q(-w_d)$ and $e^{w_d} Q_R(-w_d)$. 

\subsection{Properties in the nonequilibrium steady state}
\begin{figure}
\vspace{0.6cm}
\includegraphics[width=3.4in]{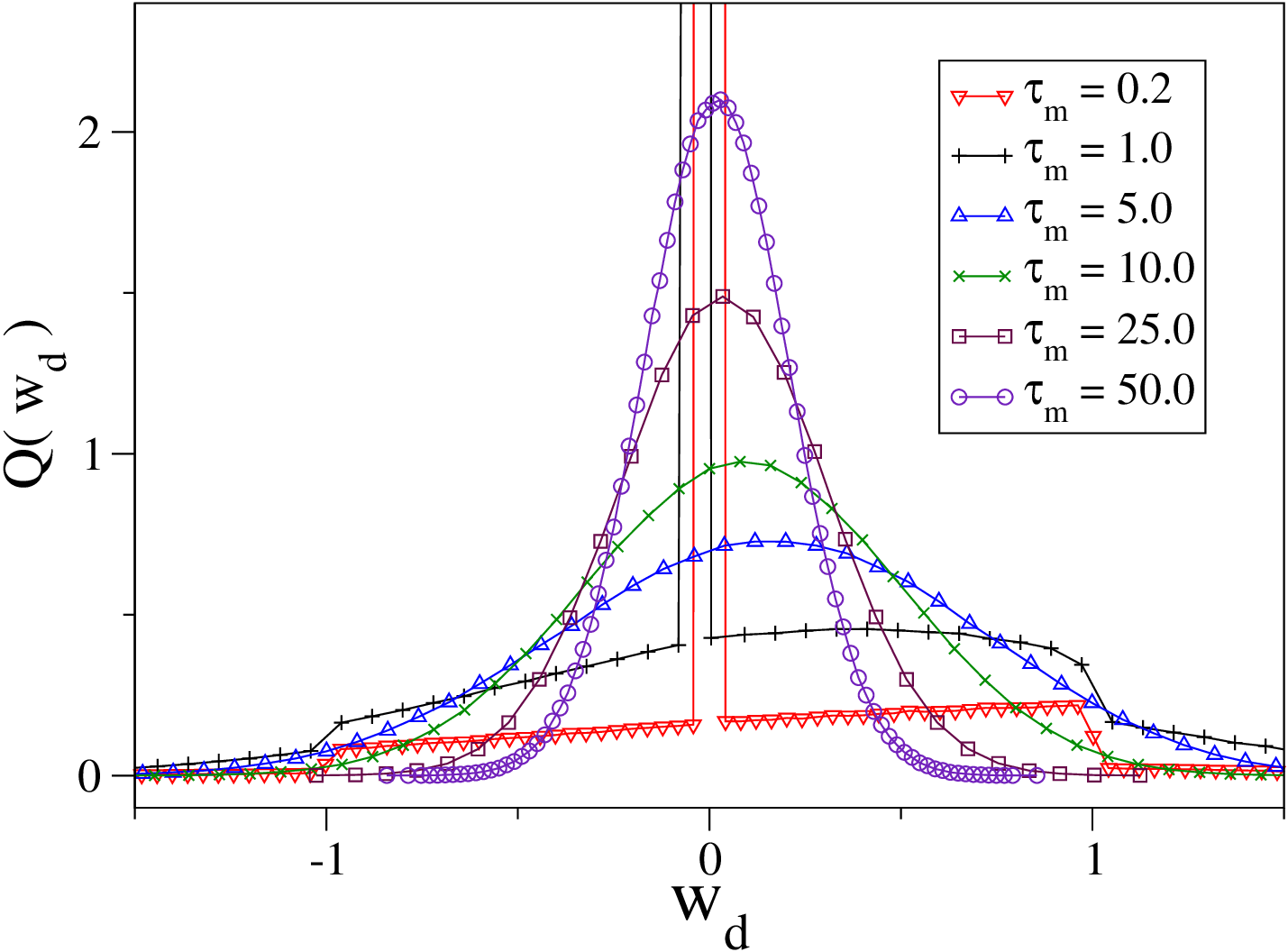}
\caption{ Work-distributions in the nonequilibrium steady state.}
\label{wdss}
\vspace{0.7cm}
\end{figure}

\begin{figure}
\vspace{0.6cm}
\includegraphics[width=3.4in]{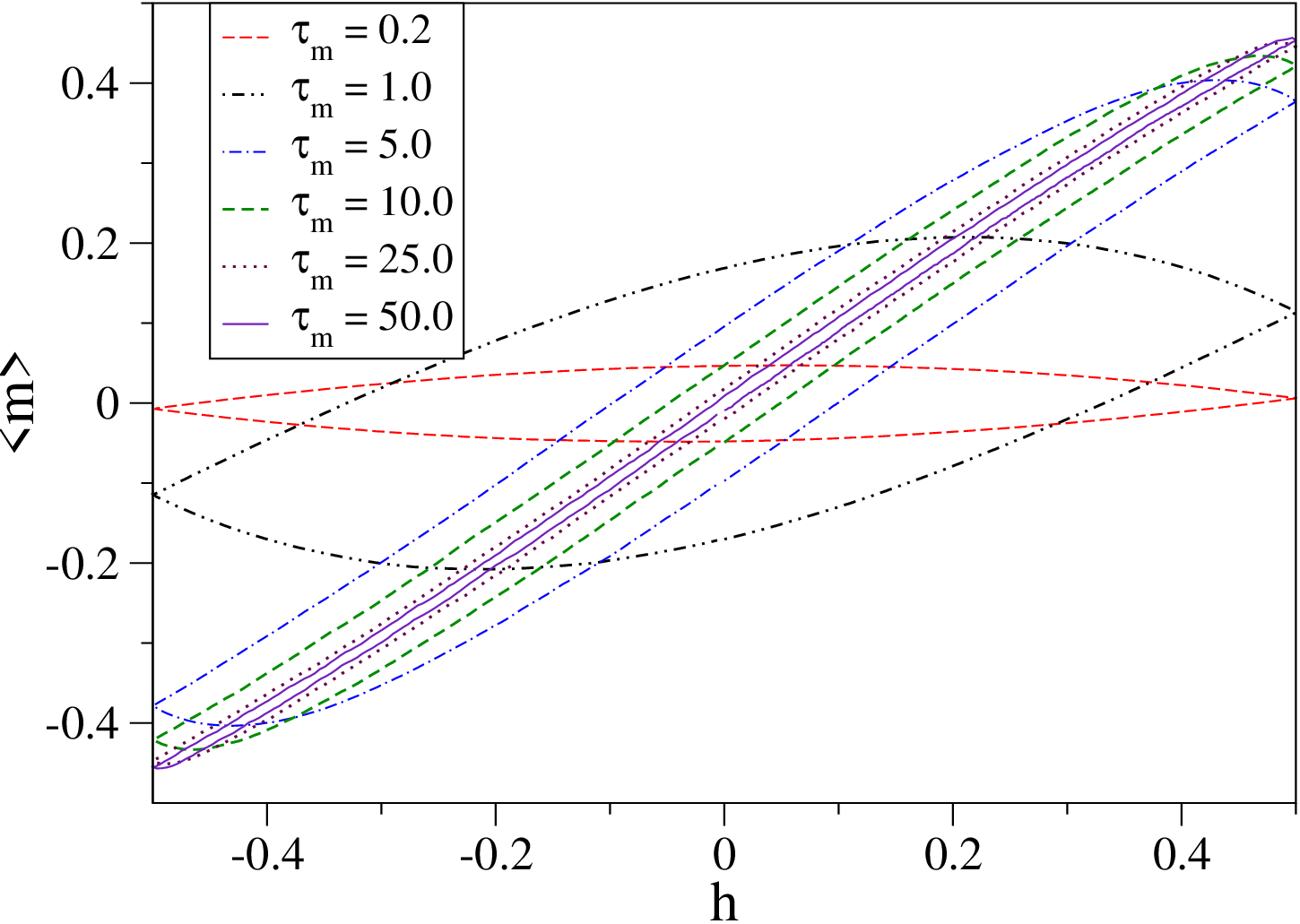}
\caption{Hysteresis curves in the nonequilibrium steady state.}
\label{magss}
\vspace{0.65cm}
\end{figure}

We now look at the case when the spin is driven by the oscillating
field into a nonequilibrium steady state and we measure fluctuations
in this steady state. In this case the work distributions (over a cycle)
have the same forms as in the transient case (Fig.~\ref{wdss}). 
The joint distribution function $Q(w,\s,\tau)$ satisfies the same
equation Eq.~\ref{eqQ} but now the initial conditions are different. 
In Fig.~\ref{magss} we plot the steady state hysteresis
curves. Note that unlike the transient case the hysteresis curves are
now closed loops.

\begin{figure}
\vspace{0.6cm}
\includegraphics[width=3.4in]{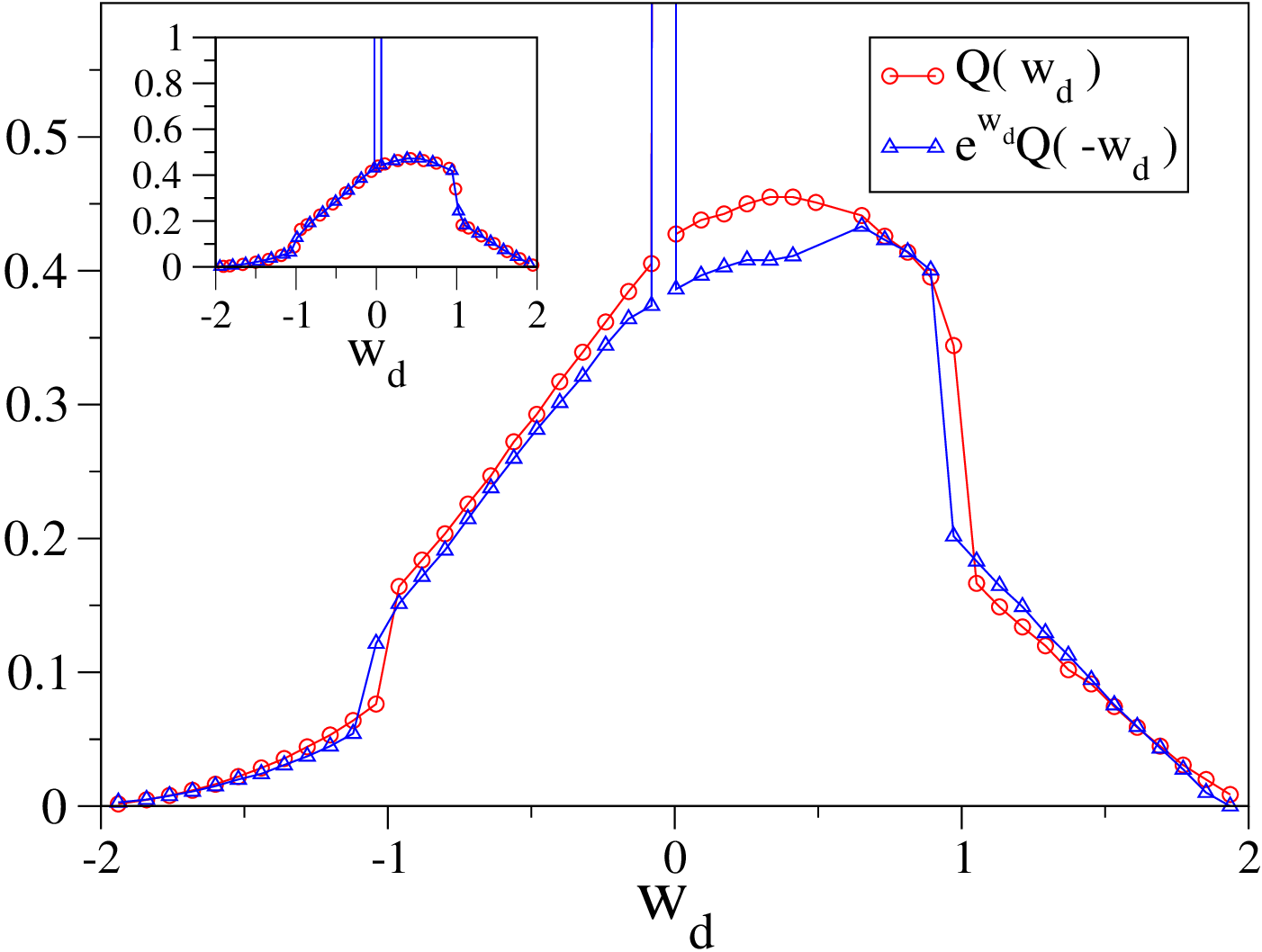}
\caption{ Violation of the  fluctuation theorem for the steady state
  work-distribution corresponding to $\t_m=0.5$. Inset shows the same plots
  for the transient case where FT is clearly satisfied.}
\label{ssft}
\vspace{0.65cm}
\end{figure}

Finally we test the validity of the steady state
fluctuation theorem (SSFT). This theorem has been proved for dynamical
systems evolving through deterministic equations but there exists no
proof that a similar result holds for stochastic dynamics.   
From Fig.~\ref{ssft} it is clear that SSFT does not hold. 

\section{Analytic results for slow and fast rates}

\subsection{Field increased linearly from $0$ to $\th$}

(i) \underline{Slow case}: $\t_m  >> 1$ 

\begin{figure}
\vspace{0.6cm}
\includegraphics[width=3.4in]{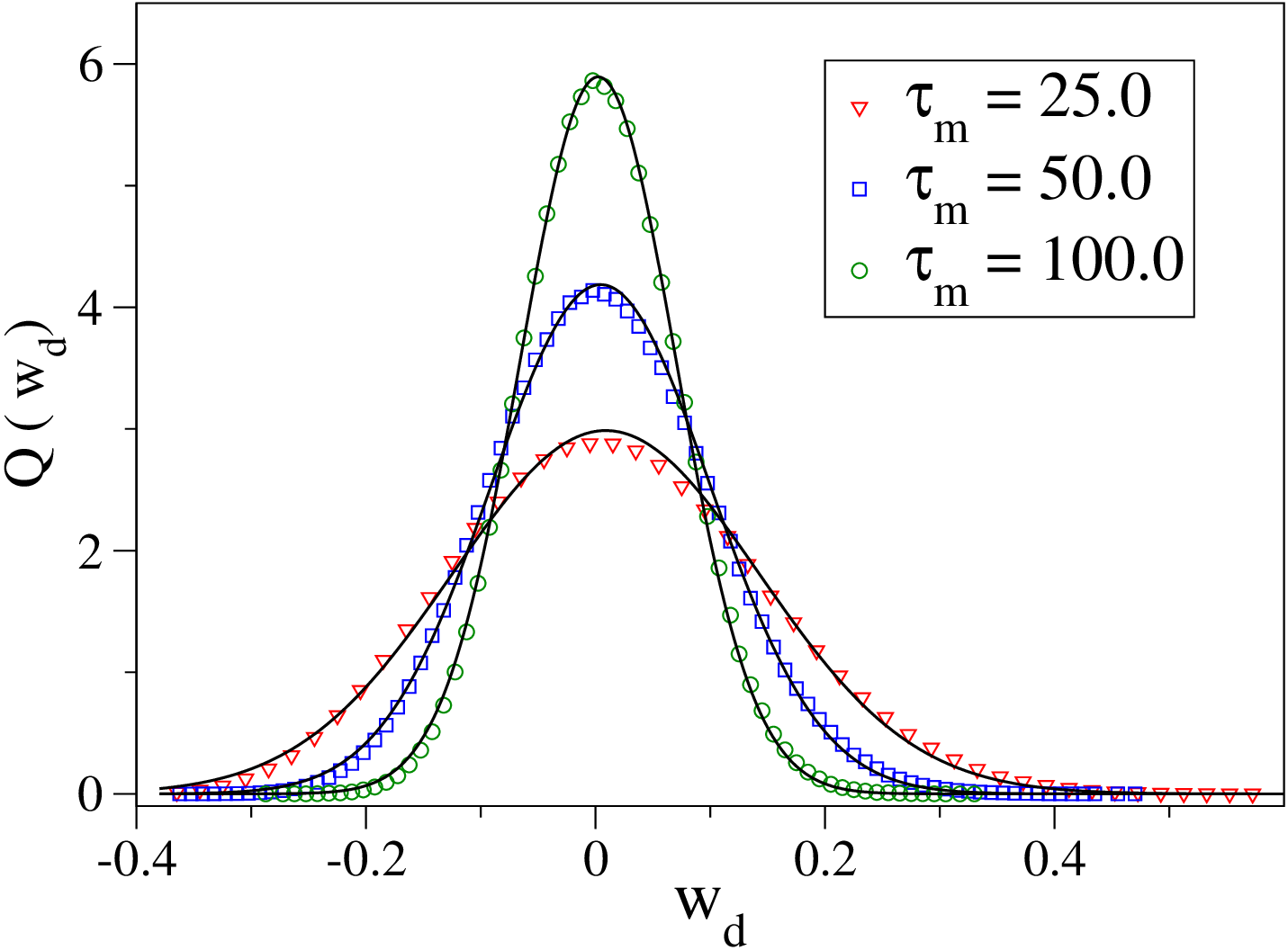}
\caption{
Comparison of work-distributions for slow rates obtained from
simulations and from the analytic form. Solid lines show the analytic results.}
\label{slowfitA}
\vspace{0.65cm}
\end{figure}

As argued in the previous section we expect that the
   work-distributions to be Gaussians which will be of  the general form
\bea
Q(w)=\frac {1}{\sqrt {2\pi \sigma^2}}\ e^{\frac{-(w-\la w \ra)^2}{2 \sigma_w^2}\ }\ .
\eea 
Since the distribution satisfies Jarzynski's equality, it follows at
once that the mean and variance are related by 
\bea
\sigma_w^2 = 2 (\la {w}\ra -{\beta \D F}).
\eea
Hence  we just need to find the mean work done. The mean work done is given
   by $\la w\ra=-(\th/\t_m) \int_0^{\t_m} d\t m(\t)$. 
 In the strict    adiabatic limit $\t_m \to \infty$ we have
   $m_{ad}(\t)=tanh(\th \t/\t_m)$ and the mean work done $\la w \ra=
   -\log ( \cosh(\th))= \beta \Delta F$. For large $\t_m$ we
   try the perturbative solution 
\bea
m(\t)=m_{ad}(\t)+\f{1}{\t_m} g(\t)
\eea
Substituting in Eq.~\ref{mageq} we get an equation for $g(\t)$ whose
   solution gives
\bea
g(\t)=-(\th)sech^2{(\f{\th \tau}{\tau_m})} + O(\f{1}{\t_m^2}) 
\eea
For the work done we then get
\bea
\la w \ra = \beta \D F+\frac{\th}{\t_m}tanh(\th)
\eea
In Fig.~\ref{slowfitA} we compare the simulations for slow rates with
   the analytic results.     

(ii) \underline{Fast case}: $\t_m << 1$.

\begin{figure}
\vspace{0.6cm}
\includegraphics[width=3.4in]{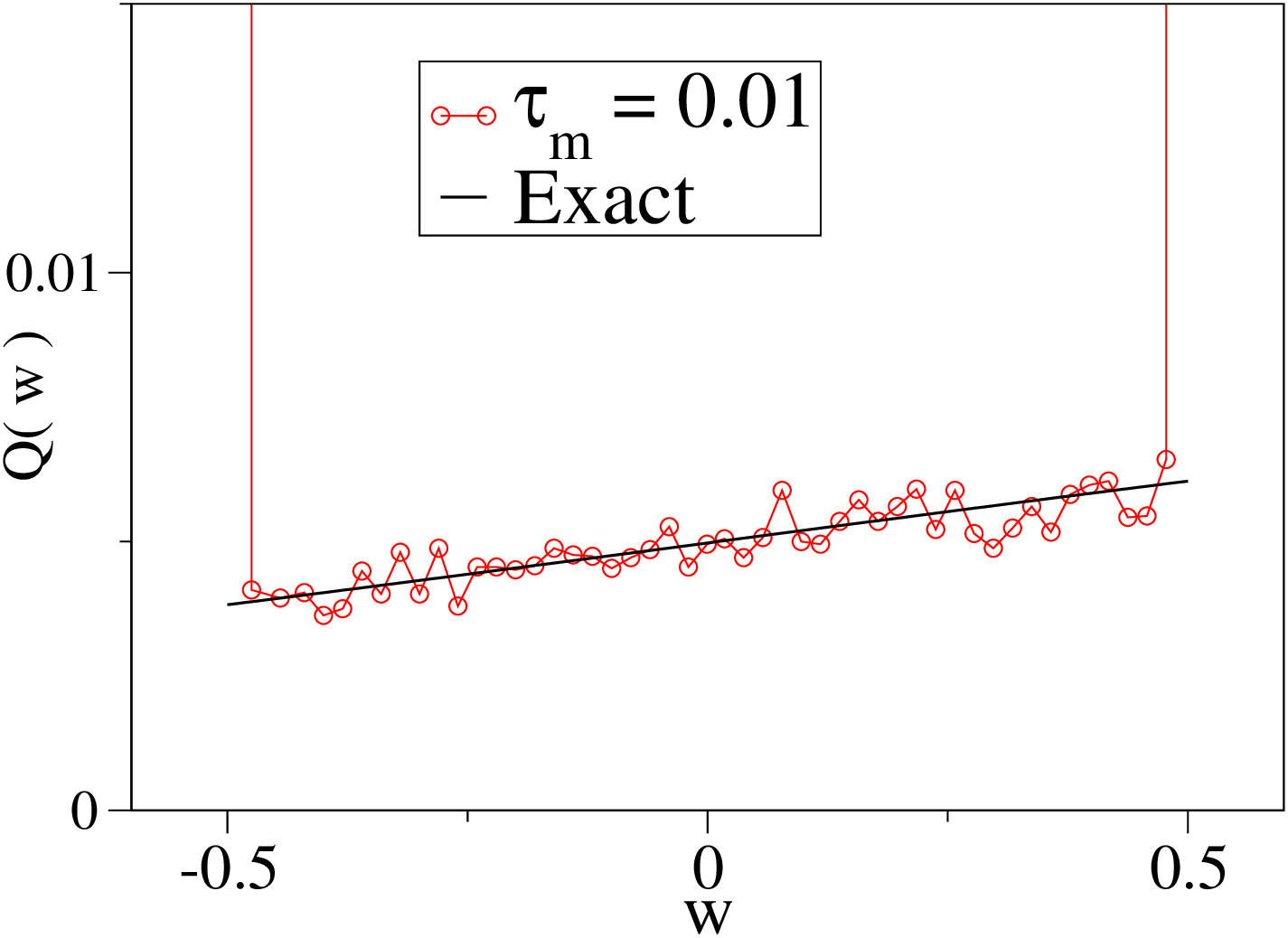}
\caption{Comparison of work-distribution for a fast rate 
  obtained from  simulation and from the analytic form.}
\label{fastfitA}
\vspace{0.65cm} 
\end{figure}

If we change the field very fast then the spin is not able to respond
    and so there are few spin flips during the entire process. At
    the lowest order there is no flip and this gives rise to the
    $\d-$functions peaks at $\pm\th$ seen in the distribution. We now
    calculate the work-distribution by looking at contributions from
    $0-$spin flip and $1-$spin flip processes. 
Let $S(\up,\t_0,\t)$ be the probability that, given that the spin is $\up$ at time
    $\t_0$, it remains in the same state till time $\t$. It is
    easy to see that $S(\up,\t_0,\t)$ satisfies the equation   
\bea 
\frac{\p S(\up,\t_0,\t)}{\p \t}\  = - \frac{e^{-\th f(\t)}}{Z}
    S(\up,\t_0,\t)
\label{nofleq}
\eea
Solving we get,  for the linear case $f(\t)=\t/\t_m$,
\bea 
S(\up,0,\t) =
    e^{-\int_{0}^{\t}\frac{e^{-\th\frac{\t^{\prime}}{\t_m}}}{Z(\t^{\prime})}
    d\t^{\prime}} =
    e^{\f{-\t}{2}}(\cosh(\f{\th\t}{\t_m}))^{\f{\t_m}{2\th}}
\eea
Putting $\t=\t_m$ corresponds to the process for which the work done
    is $w=-\th$. Hence, since the probability of the spin being
    initially in
    $\up$ state is $1/2$, we get
\bea 
Prob(w=-\th) =\frac{1}{2}
    e^{-\f{\t_m}{2}}[\cosh(\th)]^{\frac{\t_m}{2\th}} 
\label{PrA1}
\eea  
Proceeding in a similar fashion by starting with a $\dn$ spin we get 
\bea 
Prob(w=\th) =\frac{1}{2}\
    e^{\f{-\t_m}{2}}(\cosh(\th))^{-\frac{\t_m}{2\th}} 
\label{PrA2}
\eea  
Next let us consider $1-$spin flip processes which (for fast rates)
    are the major contributors to the
    part of the distribution between the two peaks. Let 
    $S_1(\up,\t) d\t$ be the probability that the spin starts
    in the $\up state$, flips once between times $\t$ to $\t+d\t$, and
    stays $\dn$ till time $\t_m$.
This is given by
\bea
S_1(\up,\t) d \t= S(\up,0,\t) (e^{-\th \t/\t_m}/Z) d \t S(\dn,
    \t, \t_m)  
\eea 
The work done during such a process is given by 
\bea 
w = -\frac{\th}{\t_m}\ (2\t-\t_m)
\eea
Similarly the case where the spin starts from a $\dn$ state gives
\bea
S_1(\dn,\t) d \t= S(\dn,0,\t) (e^{\th \t/\t_m}/Z) d \t S(\up,
    \t, \t_m)  
\eea
and the work done in this case is
\bea 
w = \frac{\th}{\t_m}\ (2\t-\t_m)
\eea
Adding this two contributions and plugging in the form of $S(\s,
    \t_0,\t)$ obtained earlier we get the following  contribution to
the  work-distribution:
 \bea 
Q_1(w)= \frac{\t_{m}}{8\th} e^{\frac{-(\t_m-w)}{2}}(\frac{e^{\frac{-\th}{2}}(\cosh(\th)^{\f{\t_m}{2\th}})}
       {\cosh(\frac{\th-w}{2})}+\frac{e^{\frac{\th}{2}}(\cosh(\th)^{\f{-\t_m}{2\th}})}{\cosh(\frac{\th+w}{2})}) \nn 
\eea
The full distribution is given by
\bea
Q(w) &=& \frac{1}{2}
    e^{-\f{\t_m}{2}}[\cosh(\th)]^{\frac{\t_m}{2\th}} \d(w+\th) \nn \\ &+& \frac{1}{2}
    e^{-\f{\t_m}{2}}[\cosh(\th)]^{-\frac{\t_m}{2\th}} \d(w-\th)+Q_1(w)
\eea
for $-\th < w < \th$, and zero elsewhere.
In Fig.~\ref{fastfitA} we show a comparison of this analytic
form with simulation results for $\t_m=0.01$. The strengths of the
$\delta-$functions at $w=\pm \th$ are accurately given by Eqs.~\ref{PrA1},~\ref{PrA2}.

\subsection{Field is taken around a cycle}

(i) \underline{Slow case}: $\t_m  >> 1$ 

We again expect a Gaussian distribution and since $\Delta F=0$ for a
cyclic process, hence the mean and variance of the distribution are
related by $\s_w^2= 2\la{w}\ra$ (see Fig.14). As before we compute the mean work
to order $1/\t_m$ and find 
\bea
\la w \ra = 
 \frac{4\th}{\t_m}tanh(\th)
\eea
\begin{figure}
\vspace{0.6cm}
\includegraphics[width=3.4in]{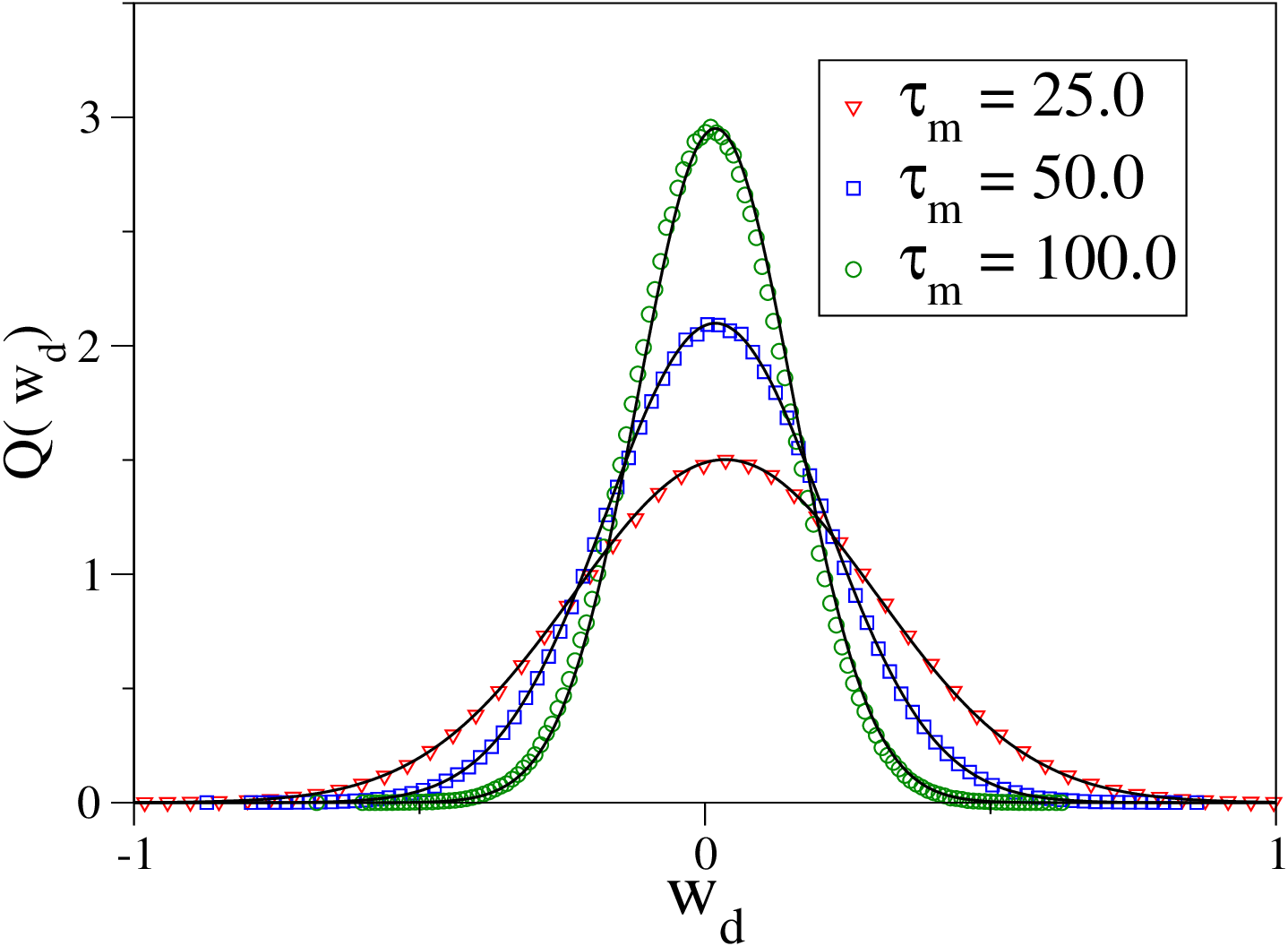}
\caption{
Comparison of work-distributions for slow rates obtained from
simulations of the cyclic case and from the analytic form. Solid lines
show the analytic results.}
\label{slowfitB}
\vspace{0.65cm}
\end{figure} 
\begin{figure}
\vspace{0.6cm}
\includegraphics[width=3.4in]{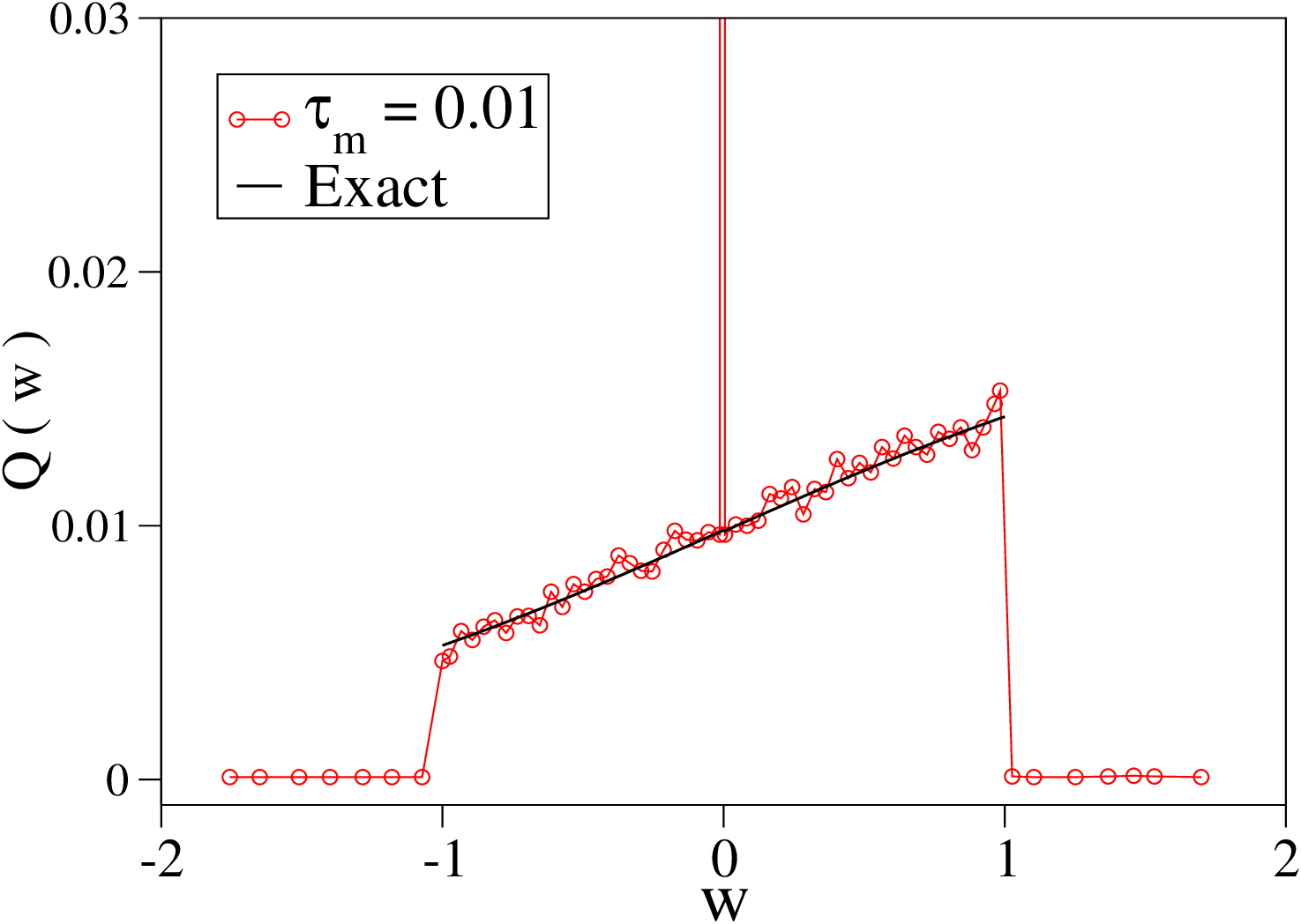}
\caption{Comparison of work-distribution, for the symmetric cycle with 
  a fast rate,  obtained from  simulations and from the analytic results.}
\label{fastfitB1} 
\vspace{0.65cm}
\end{figure}
(ii) \underline{Fast case}: $\t_m << 1$.

In this case the work distribution gives a $\d-$function peak at the
origin for $0-$spin flip processes. To find the probability of this, 
we solve Eq.~\ref{nofleq} with $f(\t)$ for the cycle given by
\bea
f(\t)&=&\f{\th}{\t_m}\t        ~~~~~~~~~~~~~     \text{for} \hspace{0.3cm}0\le\t\le\t_m \nn\\
f(\t)&=&\f{\th}{\t_m}(2\t_m -\t)  ~~\text{for} \hspace{0.3cm}\t_m\le\t\le3\t_m \nn\\
f(\t)&=&\f{\th}{\t_m}(\t - 4\t_m) ~~\text{for}
\hspace{0.3cm}3\t_m\le\t\le4\t_m \nn
\eea
This has the solution 
\bea
S(\up,0,\t_m)=e^{-2 \t_m}
\eea
Adding up an equal contribution from $S(\dn,0,\t_m)$, and since both
initial conditions occur with probability half, we finally get 
\bea
Prob(w=0)=e^{-2 \t_m}
\label{deltaCS}
\eea
Next we look at the contribution of $1-$spin flip processes. Let the
spin flip occur between times $\t$ and $\t+d\t$. It is
convenient to divide the total time $4\t_m$ into four equal intervals, 
the dependence of $w$ on $\t$ being 
different in each of the intervals. Thus if we start with the spin
initially in an $\up$ state then we have 
\bea
w&=&-\f{2 \th \t}{\t_m} ~~~~~~~~~~~~~0 <\t<\t_m \nn \\
&=&\f{2 \th}{\t_m}(\t-2 \t_m)~~~\t_m < \t < 2 \t_m \nn \\
&=&\f{2 \th}{\t_m}(2 \t_m-\t) ~~~ 2\t_m < \t < 3 \t_m \nn \\
&=&\f{2 \th}{\t_m}(\t-4 \t_m)~~~3 \t_m < \t < 4 \t_m . \nn 
\eea
The probabilities of each of these processes is
again given by:
\bea
S_1(\up,\t) d \t= S(\up,0,\t) (e^{-\th f(\t)}/Z) d \t S(\dn,
    \t, 4\t_m)  
\eea 
Using the relations between $w$ and $\t$ and summing up the four
different possibilities we finally get (for initial spin state $\up$) 
\bea
Q_1^{\up}(w)&=&\f{\t_m}{8\th}e^{-2\t_m}e^{\f{w}{2}}[(\cosh(\f{w}{2}))^{\f{\t_m}{\th}-1}
  \nn \\ &+&
(\cosh(\f{w}{2}))^{-\f{\t_m}{\th}-1}(\cosh(\th))^{\f{2\t_m}{\th}}]
\eea
for $ -1 < w < 1$ and zero elsewhere. Note that the allowed 
range of $w$ is $[-2,2]$ but single spin-flip processes only
contribute to work in the range $[-1,1]$. 
Similarly if we start with spin state $\dn$ we get
\bea
Q_1^{\dn}(w)&=&\f{\t_m}{8\th}e^{-2\t_m}e^{\f{w}{2}}[(\cosh(\f{w}{2}))^{-\f{\t_m}{\th}-1}
  \nn \\ 
  &+& (\cosh(\f{w}{2}))^{\f{\t_m}{\th}-1}(\cosh(\th))^{-\f{2\t_m}{\th}}]
\eea
for $ -1 < w < 1$.
The full work-distribution (contribution from $1-$spin flip processes)
is thus:
\bea
Q(w)=e^{-2\t_m} \d(w)+ Q_1^{\up}(w)+Q_1^{\dn}(w)
\eea
In Fig.~\ref{fastfitB1} we compare the analytic and simulation
results. The strength of the $\delta-$function at $w=0$ is accurately
given by Eq.~\ref{deltaCS}.

\section{Discussion} 
We have computed probability distributions of the work done when a
single spin, with Markovian dynamics, is driven by a time dependent
magnetic field. We find that work fluctuations are quite large (even
for slow driving rates) and there is significant probability for
processes with negative dissipated work. For slow driving the number of spin flips
during the entire process is very large and the total work is 
effectively a sum of random variables. Hence the  distributions are
Gaussian with widths proportional to the driving rate. On the other
hand for very fast driving the probability of flipping is low and we
can compute the work-distributions perturbatively from probabilities of
zero-flip, one-flip, etc. processes. 

While the two special cases of
slow and fast rates can be solved, it looks difficult to obtain a
general solution valid for all rates even in this single particle
problem. We note that the problem of 
calculating the work-distribution  is similar to that of calculating
residence-time distributions in stochastic processes
\cite{dornic,newman,satya}. In fact for the 
case in section(IIIA) the work done is proportional to the average
magnetization which is easily related to the residence time (time spin
spends in $\up$ state). For stationary stochastic processes, such as
the random walk, the residence time
distribution can be obtained exactly. However for non-stationary
processes this becomes difficult and no exact solutions are available \cite{satya}. 
In our spin-problem too it appears that the non-stationarity of the
process makes an exact solution difficult. 

For a system with $N$ spins the total work done on
the system is simply a sum of the work done on each of the spins. For the case where the
spins are non-interacting we thus get a sum of $N$ independent random
variables. For large $N$ the distribution will be a Gaussian with a
mean that scales with $N$ and variance as $N^{1/2}$. For interacting
spins the properties of the work-distribution is an open 
problem. Especially of interest is the question as to what happens
as we cross the transition temperature. Finally we note that the
large fluctuations in the area under a hysteresis curve should be
experimentally observable in nano-scale magnets.

\begin{acknowledgments}
We thank Arun Jayannavar and Madan Rao for helpful discussions. 
\end{acknowledgments}

\end{document}